\documentstyle[11pt,titlepage]{article}
\input epsf
\newcommand{\case}{\left\{\begin{array}{ll}}
\newcommand{\stopcase}{\end{array}\right.}

\newcommand{\proof}{\noindent{\bf Proof: }}
\newcommand{\qed}{\quad\rule{1.5mm}{1.5mm}}
\newcommand{\ds}{\displaystyle}

\newcommand{\la}{\lambda}
\newcommand{\be}{\beta}
\newcommand{\al}{\alpha}
\newcommand{\ep}{\epsilon}
\newcommand{\ga}{\gamma}
\newcommand{\del}{\delta}
\newcommand{\Del}{\Delta}
\newcommand{\th}{\theta}
\newcommand{\Th}{\Theta}
\newcommand{\sig}{\sigma}

\newcommand{\La}{\Lambda}

\newcommand{\om}{\omega}
\newcommand{\C}{{\bf C}}
\newcommand{\R}{{\bf R}}

\newcommand{\wuo}{W^u(z,c,\om,\ep)}
\newcommand{\wso}{W^s(z,c,\om,\ep)}

\newcommand{\atanh}{\mbox{Tanh}^{-1}}

\newcommand{\bfta}{{\bf\tilde A}}
\newcommand{\bfa}{{\bf A}}
\newcommand{\bftu}{{\bf\tilde U}}
\newcommand{\bfu}{{\bf U}}
\newcommand{\bfh}{{\bf H}}
\newcommand{\bfy}{{\bf Y}}

\newcommand{\bfe}{{\bf e}}
\newcommand{\tq}{\tilde{Q}}
\newcommand{\tp}{\tilde{P}}

\title{Stability criterion for bright solitary waves of \\
    the perturbed cubic-quintic Schr\"odinger equation
       }
       
\author{Todd Kapitula{ 
        \thanks{E-mail: kapitula@math.unm.edu} 
        \thanks{URL: http://www.math.unm.edu/$\sim$kapitula}}
   \\ Department of Mathematics and Statistics
   \\ University of New Mexico
   \\ Albuquerque, NM 87131
   \\ USA}

\date{\today }

\newtheorem{theorem}{Theorem}[section]
\newtheorem{lemma}[theorem]{Lemma}

\newtheorem{prop}[theorem]{Proposition}
\newtheorem{cor}[theorem]{Corollary}

\newtheorem{remark}[theorem]{Remark}
\newtheorem{ass}[theorem]{Assumption}

\begin{document}
\maketitle

\renewcommand{\theequation}{\arabic{section}.\arabic{equation}}
\noindent{\bf ABSTRACT.} The stability of the bright 
solitary wave solution to the perturbed
cubic-quintic Schr\"odinger equation is considered.
It is shown that in a certain region of parameter space these solutions
are unstable, with the instability being manifested as a small
positive eigenvalue.
Furthermore, it is shown that in the complimentary region of
parameter space there are no small unstable eigenvalues.
The proof involves a novel calculation of the Evans function,
which is of interest in its own right.
As a consequence of the eigenvalue calculation, it is additionally 
shown that $N$-bump bright solitary waves bifurcate from 
the primary wave.


\section{Introduction}
\setcounter{equation}{0}

The nonlinear cubic-quintic Schr\"odinger equation (CQNLS) is given by
\begin{equation}\label{eqnlcqs}
iA_t=A_{xx}+|A|^2A+\al|A|^4A,
\end{equation}
where $A$ is a complex-valued function of the variables $(x,t)\in
\R\times\R^+$.
When $\al=0$, the equation becomes the focusing
cubic nonlinear Schr\"odinger equation, and is used to
describe the propagation of the envelope of a light pulse in an optical
fiber which has a Kerr-type nonlinear refractive index.
For short pulses and high input peak pulse power the 
refractive index cannot be described by a Kerr-type nonlinearity, as
the index is then influenced by higher-order nonlinearities.
In materials with high nonlinear coefficients, such as 
semiconductors, semiconductor-doped glasses, and organic polymers, 
the saturation of the nonlinear 
refractive-index change is no longer neglible at moderately 
high intensities and should be taken into account (\cite{gatz:spi91}).
Equation (\ref{eqnlcqs}) is the correct model 
to describe the  propagation of the envelope of a light pulse 
in dispersive materials with either a saturable or higher-order
refraction index (\cite{gatz:spi91}, \cite{gatz:sca92}).

Equation (\ref{eqnlcqs}) cannot really be thought of as a 
small perturbation of the cubic nonlinear Schr\"odinger equation,
as it has been shown that a physically realistic value for the
parameter $\al$ is $|\al|\sim 0.1$ (\cite{herrmann:bbs92}).
It turns out that the most physically interesting behavior
occurs when the nonlinearity is saturating, 
so for the rest of this paper it will be assumed that $\al<0$
(\cite{angelis:stp94}, \cite{gatz:spi91}, \cite{gatz:spa92},
\cite{herrmann:bbs92}, \cite{sombra:bpc92}).
An optical fiber which satisfies this condition can be
constructed, for example, by doping with two appropriate materials
(\cite{angelis:stp94}).

One of the more physically interesting phenomena
associated with the double-doped optical fiber is
the existence of bright solitary wave solutions 
($|A(x)|\to0$ as $|x|\to\infty$) in which the peak amplitude becomes 
a two-valued function of the pulse duration.
These solutions were proven to be stable as solutions to the 
CQNLS
(\cite{boling:oso95}, 
\cite{gatz:spi91}, \cite{grillakis:sto87}, \cite{herrmann:bbs92}).

Equation (\ref{eqnlcqs}) describes an idealized fiber; therefore,
it is natural to consider the perturbed CQNLS (PCQNLS)
\begin{equation}\label{eqperturbnlcqs}
iA_t=(1+i\ep a)A_{xx}+i\ep b A+(1+i\ep d_1)|A|^2A+
    (\al+i\ep d_2)|A|^4A,
\end{equation}
where $0<\ep\ll1$ and the other parameters are real and of $O(1)$
(\cite{kodama:ssa92}).
The parameter $a$ describes spectral filtering,
$b$ describes the linear gain or loss due to the fiber, and
$d_1$ and $d_2$ describe the nonlinear gain or loss due to
the fiber.
Note that (\ref{eqperturbnlcqs}) is a well-defined PDE for $\ep>0$
only if $a>0$.

Solitary wave solutions to (\ref{eqperturbnlcqs}) are found by
setting
\begin{equation}\label{eqsolutionansatz}
A(x,t)=A(x)e^{-i\om t},
\end{equation}
and then finding heteroclinic and homoclinic solutions for the ODE
\begin{equation}\label{eqperturbnlcqsode}
(1+i\ep a)A''+(-\om+i\ep b)A+(1+i\ep d_1)|A|^2A+
    (\al+i\ep d_2)|A|^4A=0,
\end{equation}
where $'=d/dx$.
Equation (\ref{eqperturbnlcqsode}) has been extensively studied by
many authors (\cite{de:91}, \cite{duan:fdw95}, \cite{jkp:90}, 
\cite{kapitula:bba}, \cite{kapitula:sho95}, 
\cite{mn:90}, \cite{marcq:eso94}, \cite{saarloos:fps92}).
These papers have been concerned with finding various types of 
solutions, including fronts (kinks), bright solitary waves, and
dark solitary waves.
The methods employed have been both geometric (\cite{de:91}, 
\cite{duan:fdw95}, \cite{jkp:90}, \cite{kapitula:bba}, 
\cite{kapitula:sho95}) and analytic
(\cite{mn:90}, \cite{marcq:eso94}, \cite{saarloos:fps92}).

Bright solitary waves exist when there are solutions to 
(\ref{eqperturbnlcqsode}) which are homoclinic to $|A|=0$.
When $\ep=0$ and $\om>0$ the wave is given by the expression
\begin{equation}\label{eq:waveexp}
A^2(x)=\frac{4\om}{1+\sqrt{1-\be}\,\cosh(2\sqrt{\om}\,x)},\quad
\be=-\frac{16}3\al\om.
\end{equation}
Since it is being assumed that $\al<0$, a restriction on $\be$ is 
that $0\le\be<1$.
An analytic expression for the wave exists even for $\ep>0$
(\cite{marcq:eso94}, \cite{saarloos:fps92}, \cite{crespo:sot96});
however, it will not be given here.
For the purposes of this paper it is enough to know that the
wave exists for all $\ep\ge0$.

It was previously stated that the bright solitary wave is 
a stable solution to (\ref{eqnlcqs}).
However, recent numerical work by Soto-Crespo et al \cite{crespo:sot96}
suggests that this wave
becomes an unstable solution to (\ref{eqperturbnlcqs}) for
$\ep$ nonzero.
The numerics suggest that this instability arises from the presence of a 
real eigenvalue
for the linearized problem 
moving out of the origin and into the right-half of the complex
plane.
The primary purpose of this paper is to determine if this is actually the
case, and to determine possible stability/instability
mechanisms.

When discussing the stability of the bright solitary wave, one must
locate the spectrum of the operator $L$ found by linearizing
(\ref{eqperturbnlcqs}) about the wave.
The essential spectrum is easy to determine (Henry \cite{he:81}).
When $\ep=0$, it resides on the imaginary axis with
$|\mbox{Im}\,\la|\ge\om$, while for $\ep>0$ it
can be shown to be located in the left-half of the complex
plane if $a>0$ and $b<0$ (equation (\ref{eq:essspectrum})).

The location of the point spectrum is more problematic.
It is known that when $\ep=0$, zero is an eigenvalue of multiplicity
four, and there are no other point eigenvalues (Weinstein
\cite{weinstein:mso85}, \cite{weinstein:lso86}).
For $\ep\neq 0$, two of these eigenvalues will remain at the
origin, due to the spatial and rotational invariance of the PCQNLS,
while the other two will generically move and be of $O(\ep)$.
If either eigenvalue moves into the right-half plane, then the
wave will be unstable.
Unfortunately, one cannot conclude that if both eigenvalues move into
the left-half plane, then the wave is stable.
The reason is that it may be possible for eigenvalues to move
out of the essential spectrum and into the right-half plane
for $\ep\neq0$.
This topic will be the focus of a future paper.

In this paper a determination is made as to the location of the
$O(\ep)$ eigenvalues for $0<\ep\ll1$.
In order to accomplish this task, it is necessary to perform
detailed asymptotics for the Evans function, $E(\la)$, at
$\la=0$.
The Evans function is an analytic function whose zeros correspond
to eigenvalues, with the order of the zero being the
order of the eigenvalue (\cite{alexander:ati90},
\cite{pego:eis92}).
Since the null-space of $L$ is at least two-dimensional,
$E(0)=E'(0)=0$.
Thus, when expanded about $\la=0$, the Evans function satisfies
\begin{equation}\label{eq:evansexpand}
E(\la)=E''(0)\frac{\la^2}{2!}+E'''(0)\frac{\la^3}{3!}
+E^{(4)}(0)\frac{\la^4}{4!}+O(\la^5),
\end{equation}
with all the derivatives being real-valued.
It turns out to be the case that
\[
E''(0)=B_1\ep^2+O(\ep^3)
\]
(equation (\ref{eq:evanstwoexp})), while if $E''(0)=0$, then
\[
E'''(0)=B_2\ep+O(\ep^2)
\]
(equation (\ref{eq:evansthreeexp})).
Furthermore, due to a result of Weinstein
(\cite{weinstein:mso85}, \cite{weinstein:lso86}), 
$E^{(4)}(0)=O(1)$ (Corollary \ref{cor:evansfour}).
Thus, if one can determine the signs of $B_1,\,B_2$, and $E^{(4)}(0)$,
then the $O(\ep)$ eigenvalues can be approximately located.

Equation (\ref{eqperturbnlcqsode}) defines a four-dimensional ODE
phase space.
The quantity $E''(0)$ is related to the manner in which
the stable and unstable manifolds of $A=0$ intersect in this
phase space, and a calculation of this 
quantity is similiar to the calculation which leads to the 
orientation index (Alexander and Jones \cite{alexander:esa93},
\cite{alexander:esa94}).
The calculation of $E'''(0)$ is a different matter, however.
Using the ideas presented in Kapitula \cite{kapitula:ots94}, it
is shown that $E'''(0)$ has a relationship with the projection of
a certain function onto the null-space of the linear operator.
In other words,
\[
Pf=B_3E'''(0)\,A_{\cal N},
\]
where $P$ represents the projection onto the null-space,
$f$ is a particular function which measures the manner in which
the stable and unstable manifolds intersect,
$A_{\cal N}$ is a certain basis function of the null-space of $L$,
and $B_3>0$ is a constant of proportionality.
The ideas leading to the calculation of $E'''(0)$ are 
generalized in
an upcoming paper, as they are of interest in their own
right.

For the statement of the main theorems, set 
\[
\La_i=\int_{\infty}^\infty A^{i}(x)\,dx,
\]
where $A(x)$ is defined in (\ref{eq:waveexp}), and let
\[
\La_{24}=\frac{\La_2}{\La_4},\quad\La_{d_2}=-\frac{8\om}\be
         (\om\La_{24}-\frac34).
\]
Since the wave $A(x)$ depends on $\om$, so do the above constants.
Asymptotic expansions for these constants are given in Appendix A.
Note that
\[
-\frac{\partial_\om\La_{d_2}}{\partial_\om\La_{24}}>0
\]
for $0\le\be<1$ (Proposition \ref{prop:dLambda24}).

\begin{theorem}\label{thm:main} Let $0<\ep\ll1$.
Let $0\le\be<1$.
Suppose that $d_1=d_1^*$, where $d_1^*$ is given in Remark \ref{rem:d1*}.
Further suppose that $a>0$, and that
\[
d_2-\frac13\al a<0.
\]
Set 
\[
b^*=-\frac{\partial_\om\La_{d_2}}{\partial_\om\La_{24}}(d_2-\frac13\al a)<0.
\]
If $0>b^*>b$, then there is one stable real eigenvalue and one real unstable
eigenvalue, both of which are $O(\ep)$.
However, if $0>b>b^*$, then there are two stable real eigenvalues and
zero unstable eigenvalues of $O(\ep)$.
Furthermore, except for the double eigenvalue at zero, there
are no other eigenvalues which are of $O(\ep)$.
\end{theorem}

\begin{remark} If either $a<0$ or $b>0$, then the wave is unstable
due to the presence of essential spectrum in the right-half plane.
\end{remark}

\begin{remark} If $\al=d_2=0$, then the wave is unstable.
Thus, Ginzburg-Landau perturbations of the cubic NLS will 
only support unstable solitary waves.
\end{remark}

\begin{remark} Since $a>0$ and $\al<0$, the constant $b^*$ will
be negative if and only if $d_2<0$.
Thus, one can say that $d_2<0$ is a minimal stability condition.
\end{remark}

\begin{remark}\label{rem:d1*} The wave exists if $d_1=d_1^*$, where
\[
d_1^*=\frac14a-\La_{24}b-\La_{d_2}(d_2-\frac13\al a)+O(\ep)
\]
(Corollary \ref{cor:waveexist}).
\end{remark}

The constant $d_1^*$ given in the above remark depends on
$\om$, i.e., $d_1^*=d_1^*(a,b,d_2,\om)$.
Let $\om^*$ be a fixed parameter value, so that for $\om\neq
\om^*$ the wave does not exist without varying $d_1^*$.
As a consequence of Corollary \ref{lem:defb^*} and the work of
Kapitula and Maier-Paape \cite{kapitula:sdo96} one has the following
theorem concerning the existence of $N$-pulse solutions to the
PCQNLS.

\begin{theorem}\label{thm:multipulse} Let the assumptions of
Theorem \ref{thm:main} be satisfied with $d_1^*=d_1^*(a,b,d_2,\om^*)$.
For each $N\ge2$ there exists a bi-infinite sequence $\{\om^N_k\}$,
with
\[
\lim_{|k|\to\infty}\om^N_k=\om^*,
\]
such that when $\om=\om^N_k$ there is an $N$-pulse solution to
(\ref{eqperturbnlcqsode}).
If $b<b^*$, then the $N$-pulse is unstable, and there exist at
least $N$ unstable eigenvalues.
\end{theorem}

\begin{remark} The interested reader should consult \cite{kapitula:sdo96}
for a more complete description of the dynamics associated with
(\ref{eqperturbnlcqsode}).
\end{remark}

The paper is organized in the following manner.
In Section 2 the Evans function is constructed and its asymptotic 
behavior is determined.
Sections 3 and 4 are devoted to deriving expressions for the
various derivatives of the Evans function at $\la=0$.
In Section 5 the calculations are performed.
Section 6 completes the argument leading to the two main theorems
of this paper.

\section{Construction of the Evans function}
\setcounter{equation}{0}

After the transformation $A\to Ae^{-i\om t}$, the 
PCQNLS can be written in travelling wave coordinates ($z=x-ct$) as
\begin{equation}\label{eq:pcqnls}
iA_t=(1+i\ep a)A_{zz}+icA_z+(-\om+i\ep b)A+(1+i\ep d_1)|A|^2A
+(\al+i\ep d_2)|A|^4A,
\end{equation}
where $A$ is a complex-valued function of the variables
$(z,t)\in\R\times\R^+.$
Upon setting $A=A_1+iA_2$, and denoting ${\bf A}=(A_1,A_2)$, 
(\ref{eq:pcqnls}) becomes the system
\begin{equation}\label{eq:pcqnlssystem}
J\bfa_t=(I_2+\ep aJ)\bfa_{zz}+cJ\bfa_z+(-\om I_2+\ep bJ)\bfa
   +(I_2+\ep d_1J)|\bfa|^2\bfa+(\al I_2+\ep d_2J)|\bfa|^4\bfa,
\end{equation}
where $I_2$ is the $2\times2$ identity matrix and $J$ is the skew-symmetric
matrix
\[
J=\left(\begin{array}{rr}0&-1\\1&0\end{array}\right).
\]
The above system can be rewritten as
\begin{equation}\label{eq:pcqnlsabbrevsys}
J{\bf A}_t=B{\bf A}_{zz}+cJ\bfa_z+F({\bf A},\om,\ep),
\end{equation}
where 
\[
B=I_2+\ep aJ
\]
and
\[
F({\bf A},\om,\ep)=(-\om I_2+\ep bJ)\bfa
   +(I_2+\ep d_1J)|\bfa|^2\bfa+(\al I_2+\ep d_2J)|\bfa|^4\bfa.
\]

Let $\bfta$ represent the bright solitary wave solution to
(\ref{eq:pcqnlsabbrevsys}) which is known to exist when
$c=0$ (\cite{marcq:eso94}, \cite{saarloos:fps92}).
When $\ep=0,\,\bfta=(R_0,0)^T$, where
\begin{equation}\label{eq:R_0}
R_0^2(z)=\frac{4\om}{1+\sqrt{1-\be}\,\cosh(2\sqrt{\om}\,z)},\quad
\be=-\frac{16}3\al\om
\end{equation}
(\cite{crespo:sot96}).
Linearizing about the wave, the eigenvalue equation is given by
\begin{equation}\label{eq:pcqnlseval}
B{\bf A}''+DF_{{\bf A}}(\bfta,\om,\ep){\bf A}=\la J{\bf A},\quad
'=\frac{d}{dz},
\end{equation}
i.e., $-JL\bfa=\la\bfa$, where
\begin{equation}\label{eq:defL}
-JL=-J(B\partial_z^2+DF_{{\bf A}}(\bfta,\om,\ep)).
\end{equation}
A routine calculation shows that the 
essential spectrum for the
operator $-JL$, hereafter referred to as $\sig_e(-JL)$, is
given by
\begin{equation}\label{eq:essspectrum}
\sig_e(-JL)=\{\la\in\C\,:\,\la=-\ep(a\eta^2-b)\pm(\eta^2+\om)i,\,\eta\in\R\}
\end{equation}
(Henry \cite{he:81}). 
Thus, for $\ep>0$ the operator $-JL$ is sectorial if $a>0$, and the essential
spectrum is in the left-half of the complex plane if $b<0$.
This observation leads to the following assumption, which is minimal
if the solitary wave is to be stable.

\begin{ass}\label{ass:ab} The parameters $a$ and $b$ are such
that $a>0$ and $b<0$.
\end{ass}

After setting ${\bf Y}=({\bf A},{\bf A}')$, the eigenvalue equation
(\ref{eq:pcqnlseval}) can be rewritten as the first-order system
\begin{equation}\label{eq:evalsys}
{\bf Y}'=M(\la,z){\bf Y},
\end{equation}
where $M$ is the $4\times4$ block matrix
\[
M(\la,z)=\left[\begin{array}{cc}0&I_2 \\
                 B^{-1}(\la J-DF_\bfa(\bfta,\om,\ep))&0\end{array}\right].
\]
For $\la\in\Omega=\C\backslash\sig_e(L)$ 
there exist complex analytic functions ${\bf Y}^s_i(\la,z)$ and
${\bf Y}^u_i(\la,z),\,i=1,2$, which are solutions to (\ref{eq:evalsys}) and
which satisfy
\[
\begin{array}{lcll}
1.\,\,&{\ds\lim_{z\to\infty}}&{\ds|{\bf Y}^s_i(\la,z)|=0},
  \quad&\bfy^s(\la,z)=(\bfy^s_1\wedge\bfy^s_2)(\la,z)\neq0 \\
\vspace{.1mm} \\
2.\,\,&{\ds\lim_{z\to-\infty}}&{\ds|{\bf Y}^u_i(\la,z)|=0},
  \quad&\bfy^u(\la,z)=(\bfy^u_1\wedge\bfy^u_2)(\la,z)\neq0
\end{array}
\]
(\cite{alexander:ati90}).
The Evans function is given by
\begin{equation}\label{eq:evans}
E(\la)=
 ({\bf Y}^u_1\wedge{\bf Y}^u_2\wedge{\bf Y}^s_1\wedge{\bf Y}^s_2)(\la,z),
\end{equation}
and by Abel's formula is independent of $z$.
The Evans function is such that for $\la\in\Omega$ it
is zero if and only if $\la$ is an eigenvalue, with the order of
the zero being the order of the eigenvalue (\cite{alexander:ati90}).
Due to the invariances of the PCQNLS, two solutions to (\ref{eq:pcqnlseval})
when $\la=0$ are $\bfa=\bfta'$ and $\bfa=J\bfta$.
As such, one can set
\begin{equation}\label{eq:defbfy}
\begin{array}{ll}
1.\,\,&{\bf Y}^s_1(0,z)={\bf Y}^u_1(0,z)=\bftu' \\
2.\,\,&{\bf Y}^s_2(0,z)={\bf Y}^u_2(0,z)=\bftu_J,
\end{array}
\end{equation}
where
\begin{equation}\label{eq:defbftu}
\bftu=(\bfta,\bfta')^T,\quad\bftu_J=(J\bfta,J\bfta')^T.
\end{equation}
The below lemma describes the asymptotic behavior of the Evans function.

\begin{lemma}\label{lem:evanslimit} With $\bfy^u_i(0,z)$ and $\bfy^s_i(0,z)$
as described in (\ref{eq:defbfy}), if $\la\in\R$, then $E(\la)<0$ as
$\la\to\infty$.
\end{lemma}

\proof It can be assumed without loss of generality 
that $\ep=0$, as if the result is true for $\ep=0$, it will then be
true for $0\le\ep\ll1$.
Assume that $\la\in\R^+$.
 
Let $\bfy=(P,Q)^T$ in (\ref{eq:evalsys}).
Upon setting $s=\sqrt{\la}\,z$ and $Q=\sqrt{\la}\,\tq$ and letting $\la\to
\infty$, equation (\ref{eq:evalsys}) becomes the autonomous system
\[
\left(\begin{array}{c}P\\ \tq\end{array}\right)'=
\left[\begin{array}{rr}0&I_2\\J&0\end{array}\right]
\left(\begin{array}{c}P\\ \tq\end{array}\right),
\]
where $'=d/ds$.
The eigenvalues of the above matrix are given by $\ga(\pm1\pm i)$, where
$\ga=\sqrt{2}/2$, so that there exists a two-dimensional unstable
subspace and two-dimensional stable subspace, with the two-dimensional
unstable subspace being Span$\{(\ga,-\ga,1,0)^T,(\ga,\ga,0,1)^T\}$ and
the two-dimensional stable subspace being
Span$\{(\ga,-\ga,-1,0)^T,(-\ga,-\ga,0,1)^T\}$.

Let $\bfe_i\wedge\bfe_j=\bfe_{ij}$.
In $\La^2(\R^4)$ the unstable subspace is represented by the vector
\[
\begin{array}{lll}
\bfy^u(+\infty)&=&(\ga,-\ga,1,0)^T\wedge(\ga,\ga,0,1)^T \\
&=&\bfe_{12}-\ga\bfe_{13}+\ga\bfe_{14}-\ga\bfe_{23}
  -\ga\bfe_{24}+\bfe_{34},
\end{array}
\]
while the stable subspace is represented by the vector
\[
\begin{array}{lll}
\bfy^s(+\infty)&=&(\ga,-\ga,-1,0)^T\wedge(-\ga,-\ga,0,1)^T \\
  &=&-\bfe_{12}-\ga\bfe_{13}+\ga\bfe_{14}-\ga\bfe_{23}
  -\ga\bfe_{24}-\bfe_{34}.
\end{array}
\]
Note that
\begin{equation}\label{eq:evansinfinity}
\bfy^u(+\infty)\wedge\bfy^s(+\infty)=-2.
\end{equation}

When $\la=0$, for each fixed $z$ both the unstable and stable subspaces
are spanned by the vectors $(R_0'(z),0,R_0''(0),0)^T$ and 
$(0,R_0(z),0,R_0'(z))^T$.
Set
\begin{equation}\label{eq:scalelimit}
\begin{array}{lll}
\bfy^u(0)&=&{\ds\lim_{z\to-\infty}e^{-2\sqrt{\om}\,z}\bfy^u(0,z)} \\
\vspace{.1mm} \\
\bfy^s(0)&=&{\ds\lim_{z\to\infty}e^{2\sqrt{\om}\,z}\bfy^s(0,z)}.
\end{array}
\end{equation}
Using the representation for $R_0$ it can then be seen that
\[
\begin{array}{lll}
\bfy^u(0)&=&{\ds\lim_{z\to-\infty}e^{-2\sqrt{\om}\,z}(R_0'(z),0,R_0''(0),0)^T
  \wedge(0,R_0(z),0,R_0'(z))^T} \\
&=&\mu\,(\sqrt{\om},0,\om,0)^T\wedge(0,1,0,\sqrt{\om})^T \\
&=&\mu(\sqrt{\om}\bfe_{12}+\om\bfe_{14}-\om\bfe_{23}+\om^{3/2}\bfe_{34}),
\end{array}
\]
where
\[
\mu=\frac{8}{(1-\be)^{1/2}}\om,\quad \be=-\frac{16}3\al\om.
\]
A similiar calculation shows that
\[
\begin{array}{lll}
\bfy^s(0)&=&{\ds\lim_{z\to\infty}e^{2\sqrt{\om}\,z}(R_0'(z),0,R_0''(0),0)^T
  \wedge(0,R_0(z),0,R_0'(z))^T} \\
&=&\mu(-\sqrt{\om}\bfe_{12}+\om\bfe_{14}-\om\bfe_{23}-\om^{3/2}\bfe_{34}).
\end{array}
\]
Note that
\[
\bfy^u(0)\wedge\bfy^s(0)=-4\om^2\mu^2.
\]

A more complete discussion of the following argument can be found
in Alexander and Jones \cite{alexander:esa94}.
An orientation of $\R^4$ is given by a nonzero element $\eta$ of
$\Lambda^4(\R^4)$.
Two ordered bases of $\R^4,\,\{{\bf w}_1,\dots,{\bf w}_4\}$ and 
$\{{\bf u}_1,\dots,{\bf u}_4\}$, have the same orientation if
${\bf w}_1\wedge\cdots\wedge{\bf w}_4$ is a positive multiple of
${\bf u}_1\wedge\cdots\wedge{\bf u}_4$ (\cite{bishop:tao68}).
Any two such bases are related by a matrix having positive 
determinant.

The functions $\bfy^u_i(\la,z)$ and $\bfy^s_i(\la,z)$ can be used
to get a basis for $\R^4$ for each $\la$ and $z$.
In particular, in a manner similiar to (\ref{eq:scalelimit}) proper 
scalings of
\[
\lim_{z\to-\infty}\bfy^u_i(\la,z),\quad
\lim_{z\to\infty}\bfy^s_i(\la,z)
\]
determine a basis for each $\la$.
Since
\[
[\bfy^u(0)\wedge\bfy^s(0)][\bfy^u(+\infty)\wedge\bfy^s(+\infty)]>0,
\]
the matrix taking the basis at $\la=0$ to that 
at $\la=+\infty$ has positive determinant, so that the two bases
have the same orientation.
The sign of $E(\la)$ for large positive $\la$ is then determined by
equation (\ref{eq:evansinfinity}), from which the conclusion of
the lemma follows.
\qed

\begin{cor}\label{cor:evansfour} When $\ep=0$, the Evans function 
satisfies
$E(0)=E'(0)=E''(0)=E'''(0)=0$, with $E^{(4)}(0)<0$.
Furthermore, $E(\la)<0$ for $\la>0$.
\end{cor}

\proof The fact that the first three derivatives of
the Evans function at $\la=0$ are zero, with the fourth derivative
being nonzero, is a direct consequence
of the work of Weinstein (\cite{weinstein:mso85}, \cite{weinstein:lso86}).
Furthermore, it is known that when $\ep=0$ the bright solitary
wave is stable, so that there exist no positive eigenvalues 
(\cite{boling:oso95}, \cite{grillakis:sto87}); hence, the
Evans function is nonzero for $\la>0$.
The fact that the fourth derivative is negative then follows
from Lemma \ref{lem:evanslimit}.
\qed


\section{Calculation of derivatives}
\setcounter{equation}{0}

For $\ep>0$ it will be generically true that $E(0)=E'(0)=0$ with
$E''(0)\neq0$.
Since $E^{(4)}(0)<0$, by calculating $E''(0)$ one will
be able to determine the location of the zeros of $E(\la)$ which are
$O(\ep)$, and hence the location of the small eigenvalues.
When $E''(0)=0$, an eigenvalue will be passing through the origin.
A determination of $E'''(0)$ will enable one to decide whether the
eigenvalue is passing into the right-half or left-half of the
complex plane.
This section is devoted to determining these quantities, and relating
them to properties of the wave.

Time independent solutions to (\ref{eq:pcqnlsabbrevsys}) satisfy
the ODE
\begin{equation}\label{eq:pcqnlssteadystate}
B{\bf A}''+cJ\bfa'+F({\bf A},\om,\ep)=0,\quad'=\frac{d}{dz},
\end{equation}
which can be written as the first-order system
\begin{equation}\label{eq:pcqnlsode}
{\bf U}'=G({\bf U},c,\om,\ep),
\end{equation}
where ${\bf U}=({\bf U}_1,{\bf U}_2)\in\R^4$ and
\[
G({\bf U},c,\om,\ep)=\left(\begin{array}{c}{\bf U}_2 \\
    B^{-1}(-F({\bf U}_1,\om,\ep)-cJ{\bf U}_2)\end{array}\right).
\]

The bright solitary wave corresponds to a solution homoclinic to
${\bf U}=0$, and is realized as the nontrivial intersection of
the two-dimensional unstable manifold, $\wuo$, with the two-dimensional
stable manifold, $\wso$.
Due to the rotational symmetry associated with the PCQNLS, there
exists no distinguished trajectory in $\wuo\cap\wso$.
However, this rotational symmetry allows one to choose a
trajectory so that $\tilde{A}_2(0)=0$, which uniquely defines
a trajectory in the two-dimensional manifold.
Set $\bftu=(\bfta,\bfta')$, so that $\bftu\subset\wuo\cap\wso$ is a
distinguished solution.

Before continuing, the following proposition is needed.
It follows immediately upon examination of 
(\ref{eq:pcqnlsabbrevsys}).

\begin{prop}\label{prop:df} The Frechet derivative of the
nonlinearity $F$ satisfies
\begin{equation}\label{eq:dfom}
DF_\om({\bf A},\om,\ep)=-{\bf A}.
\end{equation}
\end{prop}

Since $G$ depends smoothly on the parameters, so do the manifolds.
The bright solitary wave is manifested as the nontrivial intersection
of $\wuo$ and $\wso$.
Differentiating (\ref{eq:pcqnlsode}) with respect to the parameters
$c$ and $\om$ and evaluating over the wave $\bftu$ yields the systems
\begin{equation}\label{eq:diffc}
(\partial_c W^r)'=DG_\bfu(\bfta,0,\om,\ep)\,\partial_c W^r 
    +(0,-B^{-1}J\bfta')^T
\end{equation}
and
\begin{equation}\label{eq:diffom}
(\partial_\om W^r)'=DG_\bfu(\bfta,0,\om,\ep)\,\partial_\om W^r 
    +(0,B^{-1}\bfta)^T.
\end{equation}
In these equations $r\in\{u,s\}$, the result
of Proposition \ref{prop:df} is implicitly used, and
\[
DG_\bfu(\bfta,0,\om,\ep)=\left[\begin{array}{cc}0&I_2 \\
                 -B^{-1}DF_\bfa(\bfta,\om,\ep)&0\end{array}\right].
\]
Note that a consequence of these equations is that
$\partial_c(W^u-W^s)$ and $\partial_\om(W^u-W^s)$ are solutions to the
linear system
\begin{equation}\label{eq:linsystem}
\del\bfu'=DG_\bfu(\bfta,0,\om,\ep)\,\del\bfu.
\end{equation}
If one sets $\bftu_J=(J\bfta,J\bfta')$, then the following proposition is
realized.

\begin{prop}\label{prop:linsols} Four solutions to (\ref{eq:linsystem})
are given by $\bftu',\,\bftu_J,\,\partial_c(W^u-W^s),$ and 
$\partial_\om(W^u-W^s)$; furthermore, if
\[
D_2=[\partial_c(W^u-W^s)\wedge\partial_\om(W^u-W^s)\wedge\bftu'\wedge
  \bftu_J](z,0,\om,\ep)
\]
is nonzero, then the solutions are linearly independent.
\end{prop}

\proof  It has already been seen that these four functions are solutions
to the linear system.
When $D_2\neq0$, the linear independence of the solutions follows from
the fact that $D_2$ is the Wronskian.
\qed  

\begin{remark}  By Abel's formula, $D_2$ is independent of $z$.
\end{remark}

Set
\begin{equation}\label{eq:defdelU}
\del\bfu_1=\partial_c(W^u-W^s),\,\del\bfu_2=\partial_\om(W^u-W^s),\,
\del\bfu_3=\bftu',\,\del\bfu_4=\bftu_J,
\end{equation}
so that
\[
D_2=(\del\bfu_1\wedge\del\bfu_2\wedge\del\bfu_3\wedge\del\bfu_4)(z,0,\om,\ep).
\]
Assuming that $D_2\neq0$, the functions $\del U_1$ and $\del U_2$ 
grow exponentially fast in the supremum norm as $|z|\to\infty$, while the 
functions $\del U_3$ and $\del U_4$ decay exponentially fast.

Let $\bfh:\R\to\R^4$ be a uniformly bounded measurable function.
Suppose that the solution to
\begin{equation}\label{eq:linsystemnh}
\del\bfu'=DG_\bfu(\bfta,0,\om,\ep)\,\del\bfu+\bfh
\end{equation}
is desired, and
further suppose that one wishes the solution to be bounded for either
$z\to-\infty$ or $z\to\infty$.
Denoting the solution by $\del\bfu^\pm$, with $|\del\bfu^\pm(z)|\le M<\infty$
as $z\to\pm\infty$, by following the discussion in Kapitula 
\cite{kapitula:ots94} it can be seen that
\begin{equation}\label{eq:linsystemnhsol}
\del\bfu^\pm=\frac1{D_2}(c^\pm_1(\bfh)\del\bfu_1+c^\pm_2(\bfh)\del\bfu_2+
              c_3(\bfh)\del\bfu_3+c_4(\bfh)\del\bfu_4),
\end{equation}
where
\begin{equation}\label{eq:defci}
\begin{array}{llllll}
c^\pm_1(\bfh)&=&{\ds\int_{\pm\infty}^z|\bfh\,\del\bfu_2\,\del\bfu_3\,
                      \del\bfu_4|(s)\,ds},\quad&
c^\pm_2(\bfh)&=&{\ds\int_{\pm\infty}^z|\del\bfu_1\,\bfh\,\del\bfu_3\,
                      \del\bfu_4|(s)\,ds} \\
\vspace{.1mm} \\
c_3(\bfh)&=&{\ds\int_0^z|\del\bfu_1\,\del\bfu_2\,\bfh\,\del\bfu_4|(s)\,ds}, 
\quad&
c_4(\bfh)&=&{\ds\int_0^z|\del\bfu_1\,\del\bfu_2\,\del\bfu_3\,\bfh|(s)\,ds}.
\end{array}
\end{equation}
The following lemma can now be proved.

\begin{lemma}\label{lem:D2relations} Set
\[
\bfh_1=(0,-B^{-1}J\bfta')^T,\quad\bfh_2=(0,B^{-1}\bfta)^T.
\]
Then
\[
D_2=\int_{-\infty}^\infty|\bfh_1\,\del\bfu_2\,\del\bfu_3\,\del\bfu_4|(s)\,ds
   =\int_{-\infty}^\infty|\del\bfu_1\,\bfh_2\,\del\bfu_3\,\del\bfu_4|(s)\,ds,
\]
while 
\[
0=\int_{-\infty}^\infty|\bfh_2\,\del\bfu_2\,\del\bfu_3\,\del\bfu_4|(s)\,ds
   =\int_{-\infty}^\infty|\del\bfu_1\,\bfh_1\,\del\bfu_3\,\del\bfu_4|(s)\,ds
\]
\end{lemma}

\proof Using equations (\ref{eq:diffc}) and (\ref{eq:diffom}) along 
with equation (\ref{eq:linsystemnhsol}), one can see
that
\[
\begin{array}{lll}
\partial_c W^u&=&{\ds\frac1{D_2}(c^-_1(\bfh_1)\del\bfu_1+c^-_2(\bfh_1)\del\bfu_2+
              c_3(\bfh_1)\del\bfu_3+c_4(\bfh_1)\del\bfu_4)} \\
\partial_c W^s&=&{\ds\frac1{D_2}(c^+_1(\bfh_1)\del\bfu_1+c^+_2(\bfh_1)\del\bfu_2+
              c_3(\bfh_1)\del\bfu_3+c_4(\bfh_1)\del\bfu_4)},
\end{array}
\]
and 
\[
\begin{array}{lll}
\partial_\om W^u&=&{\ds\frac1{D_2}(c^-_1(\bfh_2)\del\bfu_1+c^-_2(\bfh_2)\del\bfu_2+
              c_3(\bfh_2)\del\bfu_3+c_4(\bfh_2)\del\bfu_4)} \\
\partial_\om W^u&=&{\ds\frac1{D_2}(c^+_1(\bfh_2)\del\bfu_1+c^+_2(\bfh_2)\del\bfu_2+
              c_3(\bfh_2)\del\bfu_3+c_4(\bfh_2)\del\bfu_4)}.
\end{array}
\]
Subtracting and using the definition of $\del\bfu_1$ and $\del\bfu_2$
then yields
\[
\begin{array}{lll}
\del\bfu_1&=&{\ds\frac1{D_2}[(c^-_1(\bfh_1)-c^+_1(\bfh_1))\del\bfu_1+
        (c^-_2(\bfh_1)-c^+_2(\bfh_1))\del\bfu_2]} \\
\del\bfu_2&=&{\ds\frac1{D_2}[(c^-_1(\bfh_2)-c^+_1(\bfh_2))\del\bfu_1+
        (c^-_2(\bfh_2)-c^+_2(\bfh_2))\del\bfu_2]}.
\end{array}
\]
Using the definitions of the $c^\pm_i$'s and the fact that $\del\bfu_i$
are linearly independent functions yields the final result.
\qed

\vspace{3mm}
Define
\begin{equation}\label{eq:e^*}
e_1^*=-\del\bfu_1\wedge\del\bfu_3\wedge\del\bfu_4,\quad
e_2^*=-\del\bfu_2\wedge\del\bfu_3\wedge\del\bfu_4.
\end{equation}
The functions $e_i^*\in\La^3(\R^4)$ for $i=1,2$;
furthermore, since $\la=0$ is an isolated eigenvalue,
both of these functions satisfy an estimate of the type
\begin{equation}\label{eq:e^*decay}
|e_i^*(z)|\le Ce^{-\mu|z|},\quad i=1,2
\end{equation}
for some positive constants $C$ and $\mu$ (Kapitula \cite{kapitula:ots94}).
For a given bounded continuous function ${\bf F}:\R\to\R^2$, define
\[
<e_i^*,{\bf F}>=\int_{-\infty}^{\infty}(\bfh\wedge e_i^*)(s)\,ds,\quad i=1,2,
\]
where $\bfh=(0,{\bf F})^T$.
With the above discussion in mind, one can rewrite Lemma 
\ref{lem:D2relations} in the following manner.

\begin{cor}\label{cor:D2relations} The constant $D_2$ is given by
\[
D_2=<e_1^*,B^{-1}\bfta>=<e_2^*,B^{-1}J\bfta'>.
\]
Furthermore,
\[
0=<e_1^*,B^{-1}J\bfta'>=<e_2^*,B^{-1}\bfta>.
\]
\end{cor}

It is now possible to relate the Evans function to the structural stability
of the wave.
The proof of the first part of the below lemma is an alternate to that found in
Alexander and Jones \cite{alexander:esa94}, and may be of interest
in its own right.

\begin{lemma}\label{lem:evans} The Evans function satisfies
\[
E''(0)=2D_2,
\]
where $D_2$ is defined in Proposition \ref{prop:linsols}.
Alternatively,
\[
E''(0)=2<e_1^*,B^{-1}\bfta>=2<e_2^*,B^{-1}J\bfta'>.
\]
\end{lemma}

\proof In this proof, the dependence of functions on the variable
$z$ will be supressed.
Upon differentiating $E(\la)$ and evaluating at $\la=0$, one sees that
\[
E''(0)=2(\partial_\la(\bfy^u_1-\bfy^s_1)\wedge
       \partial_\la(\bfy^u_2-\bfy^s_2)\wedge\bfy^s_1\wedge\bfy^s_2)(0).
\]
In the above calculation the fact that $\bfy^u_i(0)=\bfy^s_i(0)$ was
implicitly used.
Since $\bfy^s_1(0)=\bftu'$ and $\bfy^s_2(0)=\bftu_J$, all that is left
to do is show the equivalence with the first two entries making up
$E''(0)$.

Differentiating (\ref{eq:evalsys}) with respect to $\lambda$ and evaluating
at $\la=0$ one sees that
\[
(\partial_\lambda {\bf Y}^r_1)'=M(0,z)\,\partial_\lambda {\bf Y}^r_1
     -(0,-B^{-1}J\bfta')^T
\]
and
\[
(\partial_\lambda {\bf Y}^r_2)'=M(0,z)\,\partial_\lambda {\bf Y}^r_2
     -(0,B^{-1}\bfta)^T,
\]
where $r\in\{u,s\}$.
In the above equation, the fact that $J^2=-I_2$ is implicitly used.
Since $M(0)=DG_\bfu(\bfta,0,\om,\ep)$, by following the proof of 
Lemma \ref{lem:D2relations} and using the definitions
of $\bfh_i$ presented therein it can be shown that
\[
\begin{array}{lll}
\partial_\la(\bfy^u_1-\bfy^s_1)&=&-{\ds\frac1{D_2}
   [(c^-_1(\bfh_1)-c^+_1(\bfh_1))\del\bfu_1+
        (c^-_2(\bfh_1)-c^+_2(\bfh_1))\del\bfu_2]} \\
\partial_\la(\bfy^u_2-\bfy^s_2)&=&-{\ds\frac1{D_2}
   [(c^-_1(\bfh_2)-c^+_1(\bfh_2))\del\bfu_1+
        (c^-_2(\bfh_2)-c^+_2(\bfh_2))\del\bfu_2]}.
\end{array}
\]
By the result of Lemma \ref{lem:D2relations} it is then seen that actually
\[
\partial_\la(\bfy^u_1-\bfy^s_1)=-\del\bfu_1,\quad
\partial_\la(\bfy^u_2-\bfy^s_2)=-\del\bfu_2.
\]
Upon substituting the above into the expression for $E''(0)$ the first
part of the lemma is proved.

The second part of the lemma follows immediately from Corollary
\ref{cor:D2relations}.
\qed

\begin{remark}\label{rem:varlavarpar} Note that a consequence of
the above argument is that
\[
\partial_\la\bfy_1^r=-\partial_cW^r,\quad
\partial_\la\bfy^r_2=-\partial_\om W^r
\]
for $r\in\{u,s\}$.
\end{remark}

\begin{remark} Using the expansion of the Evans function given in
(\ref{eq:evansexpand}) one can then write
\[
\begin{array}{lll}
E(\la)&=&{\ds<e_1^*,B^{-1}\bfta>\la^2+E'''(0)\frac{\la^3}{3!}
         +E^{(4)}(0)\frac{\la^4}{4!}+O(\la^5)} \\
\vspace{.1mm} \\
&=&{\ds<e_2^*,B^{-1}J\bfta'>\la^2+E'''(0)\frac{\la^3}{3!}
         +E^{(4)}(0)\frac{\la^4}{4!}+O(\la^5)}.
\end{array}
\]
\end{remark}

The above discussion is predicated on the assumption that $D_2\neq0$,
which is equivalent to the manifolds $W^u$ and $W^s$ intersecting
transversely.
There are instances in which this intersection will not be transverse,
in which case $D_2=0$.
In this circumstance $E''(0)=0$ (Lemma \ref{lem:evans}), so that an
eigenvalue is passing through the origin.
In order to determine the direction in which the eigenvalue is
moving through the origin, it would be helpful to know $E'''(0)$.

Suppose that $D_2=0$ due to the fact that $\del\bfu_2=0$, i.e., because
$\partial_\om W^u=\partial_\om W^s$, while $e_1^*\neq0$.
Note that this implies that $|\partial_\om W^u|\to0$ exponentially fast
as $|z|\to\infty$.
Let $\del\bftu_2$ be any solution to (\ref{eq:linsystem}) such that
\begin{equation}\label{eq:D3}
D_3=(\del\bftu_2\wedge e_1^*)(z,0,\om,\ep)
\end{equation}
is nonzero.
The above discussion concerning concerning the construction
of solutions to (\ref{eq:linsystemnh}) can be recreated 
by substituting $\del\bfu_2$ with $\del\bftu_2$ and $D_2$ with
$D_3$.
Let $\pi:\R^4\to\R^2$ be the projection operator onto the first
two components.
The following lemma can now be proven.

\begin{lemma}\label{lem:evansthird}  Suppose that $E''(0)=0$ with
$e_1^*\neq0$.
Then
\[
E'''(0)=6<e_1^*,B^{-1}J\pi(\partial_\om W^u)>.
\]
\end{lemma}

\proof In this proof the dependence of solutions on $z$ will be supressed.
Since $E''(0)=0$ and $e_1^*\neq0$ implies that 
$\del\bfu_2=0$, by the proof of Lemma \ref{lem:evans} it is
necessarily true that $\partial_\la(\bfy^u_2-\bfy^s_2)(0)=0$.
Using this fact, a tedious calculation then shows that
\[
E'''(0)=3(\partial_\la(\bfy^u_1-\bfy^s_1)\wedge
       \partial^2_\la(\bfy^u_2-\bfy^s_2)\wedge\bfy^s_1\wedge\bfy^s_2)(0).
\]

Define the projection matrix
\[
Q=\left[\begin{array}{cc} 0&0 \\
                        B^{-1}J&0\end{array}\right],
\]
and set $\bfh_3=Q\partial_\om W^u=(0,B^{-1}J\pi(\partial_\om W^u))^T$.
Note that $Q=M_\la(0,z)$.
By Remark \ref{rem:varlavarpar} 
\[
\partial_\la\bfy^s_2=\partial_\la\bfy^u_2=-\partial_\om W^u.
\]
Using this, differentiating (\ref{eq:evalsys}) twice with
respect to $\la$, and evaluating at $\la=0$ gives
\[
(\partial^2_\la\bfy^r_2)'=M(0,z)\partial^2_\la\bfy^r_2-2Q\partial_\om W^u,
\]
where $r\in\{u,s\}$.
Using the definition of $\bfh_3$, the solutions to this ODE are
\[
\begin{array}{lll}
\partial^2_\la\bfy^u_2(0)&=&-{\ds\frac2{D_3}
       (c^-_1(\bfh_3)\del\bfu_1+c^-_2(\bfh_3)\del\bftu_2+
              c_3(\bfh_3)\del\bfu_3+c_4(\bfh_3)\del\bfu_4)} \\
\vspace{.1mm} \\
\partial^2_\la\bfy^s_2(0)&=&-{\ds\frac2{D_3}
       (c^+_1(\bfh_3)\del\bfu_1+c^+_2(\bfh_3)\del\bftu_2+
              c_3(\bfh_3)\del\bfu_3+c_4(\bfh_3)\del\bfu_4)},
\end{array}
\]
where the above functions $c_i$ are such that in their definitions
$\del\bfu_2$ has been replace with $\del\bftu_2$.
Upon subtracting one gets that
\begin{equation}\label{eq:dlala}
\partial^2_\la(\bfy^u_2-\bfy^s_2)(0)=-\frac1{D_3}
         [(c^-_1(\bfh_3)-c^+_1(\bfh_3))\del\bfu_1+
          (c^-_2(\bfh_3)-c^+_2(\bfh_3))\del\bftu_2].
\end{equation}

By the previous lemma it is known that
\[
\partial_\la(\bfy^u_1-\bfy^s_1)(0)=-\del\bfu_1.
\]
After substituting the above expressions into that for $E'''(0)$
and using the definition of $D_3$ one gets that
\[
E'''(0)=6(c^-_2(\bfh_3)-c^+_2(\bfh_3)).
\]
Evaluating this expression gives the final step in the proof.
\qed

\begin{remark} Using the expansion of the Evans function given in
(\ref{eq:evansexpand}), if $E''(0)=0$, then
\[
E(\la)=<e_1^*,B^{-1}J\pi(\partial_\om W^u)>\la^3
         +E^{(4)}(0)\frac{\la^4}{4!}+O(\la^5).
\]
\end{remark}

\section{Alternative expressions for the derivatives}
\setcounter{equation}{0}

Now that expressions are known for the various derivatives of the
Evans function, it is desirable to reduce them to 
computable quantities.
This section is devoted to that task.

It will be convenient to write everything in polar coordinates, i.e., 
\[
\bfa=(r\cos\th,r\sin\th).
\]
In polar coordinates, $\bfu=T(r,\th,s,\phi)$, where
\begin{equation}\label{eq:defT}
T(r,\th,s,\phi)=\left[\begin{array}{c}
   r\cos\th\\ r\sin\th\\ rs\cos\th-r\phi\sin\th \\ 
    rs\sin\th+r\phi\cos\th
     		      \end{array}\right],
\end{equation}
with $s=r'/r$ and $\phi=\th'$.
A routine calculation shows that
\[
DT(r,\th,s,\phi)=\left[\begin{array}{cccc}
     \cos\th&-r\sin\th&0&0\\
     \sin\th&r\cos\th&0&0\\
     s\cos\th-\phi\sin\th&-r(s\sin\th+\phi\cos\th)&r\cos\th&-r\sin\th\\
     s\sin\th+\phi\cos\th&r(s\cos\th-\phi\sin\th)&r\sin\th&r\cos\th
   	 \end{array}\right],
\]
with
\[
|DT(r,\th,s,\phi)|=r^3,
\]
so that the transformation is nonsingular except at the origin.

In polar coordinates, let the manifolds be denoted by
$W^u_p$ and $W^s_p$.
In these coordinates it is a routine calculation to show that
\begin{equation}\label{eq:polarrelations}
\begin{array}{llllll}
\del\bfu_1&=&DT\,\partial_c(W^u_p-W^s_p),\quad&
\del\bfu_3&=&DT\,\partial_z W^u_p \\
\del\bfu_2&=&DT\,\partial_\om(W^u_p-W^s_p),\quad&
\del\bfu_4&=&DT\,\partial_\th W^u_p.
\end{array}
\end{equation}
When $s=0$ the manifolds can be parameterized as
\begin{equation}\label{eq:manpar}
W^u_p=(r^u(\th,\be),\th,0,\phi^u(\th,\be))^T,\quad
W^s_p=(r^s(\th,\be),\th,0,\phi^s(\th,\be))^T,
\end{equation}
where $\be=(c,\om,\ep)$.
Now, let the underlying wave be denoted by
\[
\bfta(z)=(R(z)\cos\Th(z),R(z)\sin\Th(z)).
\]
Due to the fact that the wave is even it can be assumed that
\begin{equation}\label{eq:R(0)}
R'(0)=\Th'(0)=0,
\end{equation}
while the rotational symmetry of the PCQNLS allows one to set
\begin{equation}\label{eq:Th(0)}
\Th(0)=0.
\end{equation}
Under these assumptions, when $z=0$, i.e., when $S=R'/R=0$,
\begin{equation}\label{eq:polarrelationsupdate}
\begin{array}{l}
\partial_c(W^u_p-W^s_p)=\partial_c((r^u-r^s),0,0,(\phi^u-\phi^s))^T \\
\partial_\om(W^u_p-W^s_p)=\partial_\om((r^u-r^s),0,0,(\phi^u-\phi^s))^T \\
\partial_z W^u_p=(0,0,S'(0),\Phi'(0))^T,\,
\partial_\th W^u_p=(0,1,0,0)^T.
\end{array}
\end{equation}
Combining the above with (\ref{eq:polarrelations}) allows one to prove
the following lemma.

\begin{lemma}\label{lem:evanspolar} The Evans function satisfies
\[
E''(0)=-8R^2(0)R''(0)\,\partial_cr^u(0,0,\om,\ep)\,\partial_\om\phi^u(0,0,\om,\ep).
\]
\end{lemma}

\proof First note that
\[
D_2=(|DT|\,\partial_c(W^u_p-W^s_p)\wedge\partial_\om(W^u_p-W^s_p)\wedge
       \partial_z W^u_p\wedge\partial_\th W^u_p)(0,0,\om,\ep).
\]
By (\ref{eq:polarrelationsupdate}) and the calculation for $|DT|$,
it can be seen that
\[
D_2=-R^3(0)S'(0)\left|\begin{array}{cc}
         \partial_c(r^u-r^s)&\partial_\om(r^u-r^s) \\
         \partial_c(\phi^u-\phi^s)&\partial_\om(\phi^u-\phi^s)
		      \end{array}\right|.
\]

The steady-state equations in polar coordinates are given by
equation (\ref{eq:pcqnlsodesys}).
As a consequence of Proposition \ref{prop:symmetry},
\[
r^u(0,c,\om,\ep)=r^s(0,-c,\om,\ep),\quad 
\phi^u(0,c,\om,\ep)=-\phi^s(0,-c,\om,\ep).
\]
Thus, it can be concluded that
\[
\partial_\om(r^u-r^s)(0,0,\om,\ep)=
\partial_c(\phi^u-\phi^s)(0,0,\om,\ep)=0,
\]
while
\[
\partial_c(r^u-r^s)(0,0,\om,\ep)=2\partial_cr^u(0,0,\om,\ep),\quad
\partial_\om(\phi^u-\phi^s)(0,0,\om,\ep)=2\partial_\om\phi^u(0,0,\om,\ep).
\]
Since $S'(0)=R''(0)/R(0)$, this then yields that
\[
D_2=-4R^2(0)R''(0)\,\partial_cr^u(0,0,\om,\ep)\,\partial_\om\phi^u(0,0,\om,\ep).
\]
The statement of Lemma \ref{lem:evans} then gives the result.
\qed

\vspace{3mm}
Now that a computable expression for $E''(0)$ is known, it would be
beneficial to have an expression for $E'''(0)$.
Observation of Lemma \ref{lem:evansthird} yields that one must
first better understand 
\[
e_1^*=-\del\bfu_1\wedge\del\bfu_3\wedge\del\bfu_4\in\Lambda^3(\R^4).
\]
By (\ref{eq:polarrelations}) the above quantity can be rewritten as
\begin{equation}\label{eq:delbfuT}
\begin{array}{lll}
e_1^*&=&-(DT\,\partial_c(W^u_p-W^s_p))\wedge(DT\,\partial_z W^u_p)
               \wedge(DT\,\partial_\th W^u_p) \\
          &=&-DT^{(3)}\,(\partial_c(W^u_p-W^s_p)\wedge\partial_z W^u_p
                 \wedge\partial_\th W^u_p),
\end{array}
\end{equation}
where $DT^{(3)}$ is the $4\times4$ matrix induced by $DT$ which maps
$\Lambda^3(\R^4)$ to itself.
The matrix $DT^{(3)}$ is formed by taking all the $3\times3$ minors of
$DT$, and is given by
\[
DT^{(3)}=r^2\left[\begin{array}{cccc}
        \cos\th&\sin\th&-(s\cos\th+\phi\sin\th)&-(s\sin\th-\phi\cos\th)\\
        -\sin\th&\cos\th&s\sin\th-\phi\cos\th&-(s\cos\th+\phi\sin\th)\\
        0&0&\cos\th&\sin\th\\
        0&0&-r\sin\th&r\cos\th
        	  \end{array}\right]
\]
(\cite{fl:89}, \cite{spivak:dg70}).
Thus, in order to finish the calculation of $e_1^*$, all that is
left to determine is 
\[
(e_1^*)_p=\partial_c(W^u_p-W^s_p)\wedge\partial_z W^u_p
\wedge\partial_\th W^u_p.
\]

Set
\begin{equation}\label{eq:defxi}
\begin{array}{l}
\xi_1=\partial_z W^u_p=(R',\Th',S',\Phi')^T \\
\xi^-_2=\partial_c W^u_p,\quad\xi^+_2=\partial_c W^s_p \\
\xi_3=\partial_\th W^u_p=(0,1,0,0).
\end{array}
\end{equation}
Let the vectors $\bfe_i,\,i=1,\dots4$, be the unit vectors
in $\R^4$, and define 
\[
\bfe_{ijk}=\bfe_i\wedge\bfe_j\wedge\bfe_k.
\]
The collection of vectors $\{\bfe_{123},\bfe_{124},\bfe_{134},\bfe_{234}\}$
form a basis for $\Lambda^3(\R^4)$, so that $(e_1^*)_p$ can be
written in terms of these vectors.
Now define
\begin{equation}\label{eq:defPij}
P_{ij}^\pm=\left|\begin{array}{cc}
          (\xi_1)_i & (\xi^\pm_2)_i \\
          (\xi_1)_j & (\xi^\pm_2)_j
		 \end{array}\right|,
\end{equation}
and set $\tilde{P}_{ij}=P_{ij}^--P_{ij}^+$.
Using (\ref{eq:defxi}), a routine calculation then shows that
\[
(e_1^*)_p=\tp_{13}\bfe_{123}+\tp_{14}\bfe_{124}-\tp_{34}\bfe_{234}.
\]
A consequence of the above discussion is the following lemma.

\begin{lemma}\label{lem:bfu^T} Let $M>0$ be given, and suppose
that $\tp_{ij}=O(\ep)$ for $|z|\le M$.
Then
\[
e_1^*=\left\{\begin{array}{cl}
        -R^2(\tp_{13}\bfe_{123}+(\tp_{14}+S\tp_{34})\bfe_{124}
         -R\tp_{34}\bfe_{234})+O(\ep^2),\quad&|z|\le M \\
 	O(e^{-\mu|z|})\ep,\quad&|z|\ge M,\end{array}\right.
\]
where $\mu>0$.
\end{lemma}

\proof Let $\eta_1>0$ be such that $R^2(z)=O(e^{-\eta_1|z|})$, i.e.,
$\eta_1=2\sqrt{\om}+O(\ep)$.
Since $\Phi=O(\ep)$ implies that $\Th=O(\ep)z$, one can easily see
that for $|z|\le M$
\[
DT^{(3)}=R^2\left[\begin{array}{cccc}
        1&0&-S&0\\
        0&1&0&-S\\
        0&0&1&0\\
        0&0&0&R
        	  \end{array}\right]+O(\ep),
\]
while $\|DT^{(3)}\|=O(e^{-\eta_1|z|})$ for $|z|\ge M$.
Therefore, given the assumption on the functions $\tp_{ij}$,
for $|z|\le M$,
\[
\begin{array}{lll}
e_1^*&=&DT^{(3)}(e_1^*)_p \\
{}&=&\tp_{13}\bfe_{123}+(\tp_{14}+S\tp_{34})\bfe_{124}
  -R\tp_{34}\bfe_{234}+O(\ep^2).
\end{array}
\]

Since $\tp_{ij}=O(\ep)$ for $|z|\le M$, it is necessarily true
that for $|z|\ge M,\,\tp_{ij}=O(e^{\eta_2|z|})\ep$, which then implies
that $(e_1^*)_p=O(e^{\eta_2|z|})\ep$.
Thus, for $|z|\ge M$ one sees that
\[
\begin{array}{lll}
|e_1^*|&\le&\|DT^{(3)}\||(e_1^*)_p| \\
&=&O(e^{-\eta_1|z|})O(e^{\eta_2|z|})\ep \\
&=&O(e^{(\eta_2-\eta_1)|z|})\ep.
\end{array}
\]
Setting $\mu=\eta_1-\eta_2$, the fact that $e_1^*$ approaches zero
exponentially fast (equation (\ref{eq:e^*decay})) guarantees that 
$\mu>0$.
\qed

\vspace{3mm}
Since $B=I_2+\ep aJ$, a simple calculation shows that
$B^{-1}J=J+O(\ep)$.
Thus, using the fact that $\Th=O(\ep)$ for $|z|\le M$, 
it is not difficult to see that when $\del\bfu_2=0$,
\begin{equation}\label{eq:wu_om}
B^{-1}J\pi(\partial_\om W^u)=\left\{\begin{array}{ll}
         (0,\partial_\om R_0)^T+O(\ep),\quad&|z|\le M \\
         O(e^{-\eta_3|z|}),
               \quad&|z|\ge M,\end{array}\right.
\end{equation}
where $\eta_3>0$.

\begin{lemma}\label{lem:evansthirdeval} Let $M>0$ be given, and
suppose that $\tp_{ij}=O(\ep)$ for $|z|\le M$.
When $E''(0)=0$, the third derivative of the Evans function satisfies
\[
E'''(0)=6\int_{-\infty}^\infty R_0^2(s)\partial_\om R_0(s)\tp_{13}(s)\,ds
   +O(e^{-\eta_4 M})\ep+O(\ep^2),
\]
where $\eta_4>0$.
\end{lemma}

\proof  The integrand associated with $E'''(0)$ is given by
$\bfh_3\wedge e_1^*$, where
\[
\bfh_3=(0,B^{-1}J\pi(\partial_\om W^u))^T.
\]
Using (\ref{eq:wu_om}) and Lemma \ref{lem:bfu^T}, one
then sees that for $|z|\le M$
\[
\begin{array}{lll}
\bfh_3\wedge e_1^*&=&-R_0^2\partial_\om R_0\tp_{13}\,
                              \bfe_{4123}+O(\ep^2) \\
                       &=&R_0^2\partial_\om R_0\tp_{13}+O(\ep^2),
\end{array}
\]
while for $|z|\ge M$
\[
\bfh_3\wedge e_1^*=O(e^{-\eta_4 M|z|})\ep.
\]
In the above calculation, the fact that 
$\bfe_{4123}=\bfe_4\wedge\bfe_{123}=-1$ is used.
\qed

\begin{remark} A similiar calculation leads to the conclusion that
\[
E''(0)\approx2\int_{-\infty}^\infty R_0(s)(\tp_{14}(s)+S_0(s)\tp_{34}(s))\,ds.
\]
\end{remark}


\section{Asymptotics}
\setcounter{equation}{0}

Now that an expression for $E''(0)$ has been derived, in order
to determine the location of the eigenvalues near zero the
expressions $\partial_cr^u$ and $\partial_\om\phi^u$ must
be calculated.
In addition, in order to calculate $E'''(0)$, one must determine
$\tp_{ij}$ and show that the quantities are $O(\ep)$ for $|z|\le M$.

Set
\[
\bfa(z)=(r(z)\cos\th(z),r(z)\sin\th(z)).
\]
Let the known underlying solitary wave be denoted by
$(R,\Th,S,\Phi)$.
Note that
\begin{equation}\label{eq:Sexpress}
S=R'/R=\frac{d}{dz}\ln R.
\end{equation}
Recall the analytic expression for the wave given in 
(\ref{eq:R_0}) when $\ep=0$, i.e.,
\begin{equation}\label{eq:waveexpress}
\begin{array}{lll}
R_0^2(z)&=&{\ds\frac{4\om}
            {1+\sqrt{1-\be}\,\cosh(2\sqrt{\om}\,z)},\quad
             \be=-\frac{16}3\al\om} \\
\Th_0(z)&=&0.
\end{array}
\end{equation}
This wave will henceforth be denoted by $(R_0,0,S_0,0)^T$.
By defining
\[
\Del=1+\ep^2a^2,
\]
the steady-state ODE is
\begin{equation}\label{eq:pcqnlsodesys}
\begin{array}{rll}
r'&=&rs \\
\th'&=&\phi \\
\Del s'&=&-\Del s^2+\Del\phi^2-c(\ep as-\phi)-(-\om+\ep^2ab) \\
&{}&\quad\quad-(1+\ep^2ad_1)r^2-(\al+\ep^2ad_2)r^4 \\
\Del\phi'&=&-2\Del s\phi-c(s+\ep a\phi)-\ep[(b+a\om)+(d_1-a)r^2+(d_2-a\al)r^4].
\end{array}
\end{equation}
It should be noted that the equation for $\th$ is superfluous,
and hence is usually ignored; however, it is included here
for completeness.
After dropping the $O(\ep^2)$ terms the variational equations are given by
\begin{equation}\label{eq:varode}
\begin{array}{lll}
\del r'&=&S\del r+R\del s \\
\del\th'&=&\del\phi \\
\del s'&=&-2R(1+2\al R^2)\del r-2S\del s+
     2\Phi\del\phi+\del\om-(\ep aS-\Phi)\del c \\
{}&{}&\quad-2\ep a[(b+a\om)+(d_1-a)R^2+(d_2-a\al)R^4]\del\ep \\
\del\phi'&=&-2\ep R[(d_1-a)+2(d_2-a\al)R^2]\del r-2\Phi\del s
  -2S\del\phi \\
{}&{}&\quad-\ep a\del\om-(\ep a\Phi+S)\del c-
    [(b+a\om)+(d_1-a)R^2+(d_2-a\al)R^4]\del\ep \\
\del\om'&=&0 \\
\del c'&=&0 \\
\del\ep'&=&0.
\end{array}
\end{equation}
An observation yields the following proposition.

\begin{prop}\label{prop:symmetry} Equation (\ref{eq:pcqnlsodesys}) is
invariant under
\[
(z,c,\om,r,\th,s,\phi)\to(-z,-c,\om,r,\th,-s,-\phi).
\]
\end{prop}

An expression for $\partial_\om\phi^u(0)$ will first be determined.
Let
\begin{equation}\label{eq:phi_epdef}
\phi_\ep(z)=\del\phi(\partial_\ep W^u_p(z,c,\om)).
\end{equation}
Since
\[
\del c(\partial_\ep W^u_p(z,c,\om))=
        \del\om(\partial_\ep W^u_p(z,c,\om))=0,
\]
by using (\ref{eq:varode}) it can be seen that when $\ep=0$
\begin{equation}\label{eq:phi_epeq}
\phi_\ep'=-2S_0\phi_\ep-[(b+a\om)+(d_1-a)R_0^2+(d_2-a\al)R_0^4).
\end{equation}
By definition $\phi_\ep$ is uniformly bounded as $z\to-\infty$, so that
upon using (\ref{eq:Sexpress}) the solution to (\ref{eq:phi_epeq}) can
be written as
\begin{equation}\label{eq:phi_epsol}
\begin{array}{lll}
R_0^2(z)\phi_\ep(z)&=&{\ds-[(b+a\om)\int_{-\infty}^zR_0^2(s)\,ds
  +(d_1-a)\int_{-\infty}^zR_0^4(s)\,ds} \\
\vspace{.1mm} \\
&&\quad\quad{\ds+(d_2-a\al)\int_{-\infty}^zR_0^6(s)\,ds]}.
\end{array}
\end{equation}

By definition the function $\phi_\ep$ describes, up to $O(\ep)$,
the location of
the $\phi$-component of $W^u_p(z,c,\om)$, so that 
\begin{equation}\label{eq:phi^u}
\phi^u(0)=\ep\phi_\ep(0)+O(\ep^2).
\end{equation}
Thus, when performing calculations on $\phi^u(0)$, for $\ep>0$ small
enough it is
sufficient to perform them on $\phi_\ep(0)$.
Given (\ref{eq:phi_epsol}) and the fact that an exact expression exists 
for $R_0^2(z)$, this then
implies that rather detailed information can be gathered
regarding the variation of $\phi^u(0)$ with respect to $\om$ for
$\ep$ sufficiently small.
Set
\begin{equation}\label{eq:defLambda}
{\ds\Lambda_{m}={\int_{-\infty}^\infty R_0^m(s)\,ds},
\quad\Lambda'_{2}={\int_{-\infty}^\infty(R_0')^2(s)\,ds}.}
\end{equation}

\begin{lemma}\label{lem:phi^u} The function $\phi^u(0)$ is
given by
\[
\phi^u(0)=\ep\phi_\ep(0)+O(\ep^2),
\]
where
\[
2R_0^2(0)\phi_\ep(0)=\La_2'a-\La_2b-\La_4d_1-\La_6d_2.
\]
\end{lemma}

\proof Since $R_0$ is an even
function,
\[
\int_{-\infty}^0 R_0^m(s)\,ds=\frac12\int_{-\infty}^\infty R_0^m(s)\,ds
\]
for any positive integer $m$.
Thus, when (\ref{eq:phi_epsol}) is evaluated at $z=0$,
\[
2R_0^2(0)\phi_\ep(0)=(-\om\La_2+\La_4+\al\La_6)a-\La_2b-\La_4d_1-\La_6d_2.
\]
The function $R_0$ satisfies
\[
R_0''-\om R_0+R_0^3+\al R_0^5=0.
\]
Upon multiplying the above equation by $R_0$ and integrating by parts
one sees that
\[
\La_2'=-\om\La_2+\La_4+\al\La_6,
\]
from which the conclusion of the lemma follows.
\qed

\begin{remark} The expressions $\La_2$ and $\La_4$ are evaluated
in Appendix A.
\end{remark}

It is known that the wave exists for all $\ep>0$, with the perturbation
being regular (\cite{marcq:eso94}, \cite{saarloos:fps92}).
A necessary condition for the existence of the bright solitary wave
is that $\phi_\ep(z)$ remains uniformly bounded as $z\to\infty$.
Since $R_0(z)\to0$ as $z\to\infty$, this then yields the next
lemma.

\begin{lemma}\label{lem:waveexist} A necessary condition for
the existence of the bright solitary wave is that
\[
\La_2'a-\La_2b-\La_4d_1-\La_6d_2=0.
\]
\end{lemma}

\proof Evaluating
(\ref{eq:phi_epsol}) at $z=\infty$ and requiring that the right-hand
side be zero at the limit yields
\[
\begin{array}{lll}
0&=&(-\om\La_2+\La_4+\al\La_6)a-\La_2b-\La_4d_1-\La_6d_2 \\
{}&=&\La_2'a-\La_2b-\La_4d_1-\La_6d_2.\qed
\end{array}
\]

\begin{remark}\label{rem:phi(0)} An examination of Lemmas
\ref{lem:phi^u} and \ref{lem:waveexist}
shows that the necessary condition for the existence of the
wave implies that
\[
\phi_\ep(0)=0.
\]
\end{remark}

The expression present in the above lemma can clearly be solved
for $d_1$ in terms of the other parameters.
Before doing so, however, it will be desirable to simplify the
above expression.
As the following proposition illustrates, there is a simple relationship
between the above quantities.

\begin{prop}\label{prop:Lambda} The relations
\[
\begin{array}{llll}
1.\quad&\Lambda_6&=&{\ds\frac3{2\al}(\om\Lambda_2-\frac34\Lambda_4)} \\
\vspace{.1mm} \\
2.\quad&\Lambda_2'&=&{\ds\frac12\om\Lambda_2-\frac18\Lambda_4}
\end{array}
\]
hold true.
\end{prop}

\proof As mentioned in the proof of Lemma \ref{lem:phi^u}, the
function $R_0$ satisfies the ODE
\[
R_0''-\om R_0+R_0^3+\al R_0^5=0.
\]
Multiplying by $R_0$ and integrating by parts yields that
\[
-\La_2'-\om\La_2+\La_4+\al\La_6=0,
\]
while multiplying by $R_0'$ and integrating yields
\[
\La_2'-\om\La_2+\frac12\La_4+\frac13\al\La_6=0.
\]
Upon subtracting the above two equations one sees that
\[
\La_2'-\frac14\La_4-\frac13\al\La_6=0.
\]
The conclusion of the first part of the proposition is now clear.

The proof for the second part follows in a similiar manner.
Simply add the two equations to get the relation
\[
\om\La_2-\frac34\La_4-\frac23\al\La_6=0,
\]
from which one immediately gets the second part of the proposition.
\qed

\vspace{3mm}
Note that for $\be=-16\al\om/3$ the relation for $\La_6$ can
be rewritten as
\[
\La_6=-\frac{8\om}\be(\om\La_2-\frac34\La_4).
\]
With this observation, define
\begin{equation}\label{eq:Lambda_b}
\Lambda_{24}={\ds\frac{\Lambda_2}{\Lambda_4}},\quad 
\Lambda_{d_2}=-{\ds\frac{8\om}{\be}(\om\Lambda_{24}-\frac34)}.
\end{equation}
Note that as a consequence of Proposition \ref{prop:Lambda},
$\La_{d_2}=\La_6/\La_4$.

\begin{cor}\label{cor:waveexist} In order for the wave to exist
the parameter $d_1$ must equal $d_1^*$, where up to $O(\ep)$
\[
d_1^*={\ds\frac14a-\Lambda_{24}b-\Lambda_{d_2}(d_2-\frac13\al a)}.
\]
\end{cor}

\proof By Lemma \ref{lem:waveexist}, in order for the wave to exist it must
be true that
\[
d_1={\ds\frac{\Lambda_2'}{\Lambda_4}a-\frac{\Lambda_2}{\Lambda_4}b
  -\frac{\Lambda_6}{\Lambda_4}d_2}.
\]
The conclusion of the corollary follows after one uses the relationships
described in Proposition \ref{prop:Lambda}.
\qed

\vspace{3mm}
Now that an expression for the function $\phi^u(0)$ is known
(Lemma \ref{lem:phi^u}), it is
possible to understand its behavior when the parameter $\om$ is
varied.
A consequence of Lemma \ref{lem:phi^u} is that it is sufficient
to understand the manner in which $\phi_\ep(0)$ varies.
Since $\phi_\ep(0)=0$ when $d=d_1^*$,
a simple application of the implicit function theorem yields that
\begin{equation}\label{eq:G_om}
\partial_\om\phi_\ep(0)+\partial_{d_1}\phi_\ep(0)\partial_\om d_1^*=0.
\end{equation}
The quantities $\partial_{d_1}\phi_\ep(0)$ and 
$\partial_\om d_1^*$ are accessible, so
that the term $\partial_\om\phi_\ep(0)$ can be calculated.

\begin{lemma}\label{lem:phi_om} When $d_1=d_1^*$,
\[
\partial_\om\phi_\ep(0)=-\frac{\Lambda_4}{2R_0^2(0)}
       \left(\partial_\om\Lambda_{24} b+
  \partial_\om\Lambda_{d_2}(d_2-\frac13\al a)\right).
\]
\end{lemma}

\proof First, an examination of Lemma \ref{lem:phi^u} shows that
\[
\partial_{d_1}\phi_\ep(0)=-\frac{\Lambda_4}{2R_0^2(0)}.
\]
Upon differentiating the expression for $d_1^*$ given in 
Corollary \ref{cor:waveexist} and using (\ref{eq:G_om}), one 
then arrives at the conclusion of the lemma.
\qed

\vspace{3mm}
It is important to understand how $\partial_\om\phi_\ep(0)$ varies with
the parameters.
Before making a definitive statement, the following proposition is
needed.

\begin{prop}\label{prop:dLambda24} Set
\[
\be=-\frac{16}{3}\al\om.
\]
When $0\le\be<1$,
\[
\partial_\om\Lambda_{24}<0,\quad\partial_\om\Lambda_{d_2}>0,
\]
so that
\[
\frac{\partial_\om\Lambda_{d_2}}{\partial_\om\Lambda_{24}}<0.
\]
\end{prop}

\proof Using the definition of $\Lambda_{24}$ and Corollary
\ref{cor:Lambdaderiv} one sees that
\[
\partial_\om\Lambda_{24}=\frac{(\Lambda_4-4\om\Lambda_2)\partial_\om\Lambda_2}
  {\Lambda_4^2}.
\]
Upon using the Taylor expansions given in Corollary \ref{cor:LambdaTaylor}
one sees that
\[
\Lambda_4-4\om\Lambda_2=-16\om^{3/2}\sum_{n=0}^\infty
   \frac{\be^n}{(2n+1)(2n+3)},
\]
which is clearly negative.
Since Corollary \ref{cor:Lambdaderiv} states that $\partial_\om\Lambda_2>0$
for $0\le\be<1$, it is now clear that $\partial_\om\Lambda_{24}<0$.

Since $\partial_\om\be=\be/\om$, 
\[
\partial_\om\La_{d_2}=-\frac{8\om}{\be}\partial_\om(\om\La_{24}).
\]
Using the Taylor series expansions given in Corollary \ref{cor:LambdaTaylor},
after some tedious manipulations one can see that
\[
\partial_\om(\om\La_{24})=C\sum_{n=0}^\infty(a_n-b_n)\be^n,
\]
where
\[
C=\left(4\om(\sum_{n=0}^\infty\frac1{2n+3}\be^n)^2\right)^{-1}>0
\]
and
\[
a_n=\sum_{j=0}^n\frac{j}{2j+1}\frac1{2(n-j)+3},\quad
b_n=\sum_{j=0}^n\frac{j}{2j+3}\frac1{2(n-j)+1}.
\]

The claim regarding $\partial_\om\La_{d_2}$ will be proven as
soon as it can be shown that $a_n-b_n<0$.
Upon combining terms,
\[
a_n-b_n=4\sum_{j=0}^nj\frac{n-2j}{f(j,n)},
\]
where
\[
f(j,n)=(2j+1)(2(n-j)+1)(2j+3)(2(n-j)+3).
\]
By the integral test,
\[
a_n-b_n\le4\int_0^nxg(x,n)\,dx,
\]
where
\[
g(x,n)=\frac{n-2x}{f(x,n)}.
\]
Set $y=x-n/2$.
Then
\[
g(y,n)=-2\frac{y}{f(y,n)},
\]
with
\[
f(y,n)=\frac12(4y^2-(n+1)^2)(4y^2-(n+3)^2),
\]
so that $g(y,n)$ is odd in $y$ with $yg(y,n)<0$.
Therefore,
\[
\begin{array}{lll}
{\ds\int_0^nxg(x,n)\,dx}&=&{\ds\int_{-n/2}^{n/2}(y+\frac{n}2)g(y,n)\,dy} \\
\vspace{.1mm} \\
&=&{\ds\int_{-n/2}^{n/2}yg(y,n)\,dy} \\
\vspace{.1mm} \\
&<&0,
\end{array}
\]
so that $a_n-b_n<0$.
\qed

\vspace{3mm}
Combining the above results yields the following corollary, which
concerns the variation of $\phi^u(0)$ with $\om$.

\begin{cor}\label{lem:defb^*} Suppose that $d=d_1^*$.
Set
\[
b^*={\ds-\frac{\partial_\om\Lambda_{d_2}}{\partial_\om\Lambda_{24}}
        (d_2-\frac13\al a)}.
\]
For $\ep>0$ sufficiently small, if $b>b^*$, then 
$\partial_\om\phi^u(0)>0$; otherwise,
$\partial_\om\phi^u(0)<0$.
Furthermore, for $0\le\be<1$
\[
\frac{\partial_\om\Lambda_{d_2}}{\partial_\om\Lambda_{24}}<0.
\]
\end{cor}

\proof Since $\La_4/R_0^2(0)>0$ and $\partial_\om\La_{24}<0$, the
result follows immediately from Lemma \ref{lem:phi_om} and 
Proposition \ref{prop:dLambda24}.
\qed

\vspace{3mm}
Now that $\partial_\om\phi^u(0)$ is known, the quantities $\partial_cr^u(0)$ 
and $\tp_{ij}$ must be calculated.
This can be accomplished simultaneously.
As in (\ref{eq:defxi}), set
\[
\xi_1=\partial_zW^u_p,\quad
\xi_2^-=\partial_cW^u_p,\quad\xi_2^+=\partial_cW^s_p.
\]
Letting $P_{x_ix_j}$ denote $\del x_i\wedge\del x_j$, as in 
(\ref{eq:defPij}) set
\[
P_{rs}^\pm=P_{rs}(\xi_1,\xi_2^\pm),\quad
P_{r\phi}^\pm=P_{r\phi}(\xi_1,\xi_2^\pm),\quad
P_{s\phi}^\pm=P_{s\phi}(\xi_1,\xi_2^\pm).
\]
Note that the computation of $E'''(0)$ requires that
$P_{rs}^--P_{rs}^+$ be known (Lemma \ref{lem:evansthirdeval}).
Before continuing, a preliminary lemma is needed.

\begin{lemma}\label{lem:phi_c} Set
\[
\phi_c^\pm=\del\phi(\xi_2^\pm).
\]
When $\ep=0,\,\phi_c^\pm=-1/2$.
\end{lemma}

\proof It is easy to see from the variational equation (\ref{eq:varode})
that when $\ep=0$
\begin{equation}\label{eq:phi_c}
(\phi_c^\pm)'=-2S_0\phi_c^\pm-S_0.
\end{equation}
This equation is easily solved, and one then finds that
\begin{equation}\label{eq:phi_csol}
\begin{array}{lll}
R_0^2(z)\phi_c^\pm(z)&=&-{\ds\int_{\pm\infty}^zR_0^2(s)S_0(s)\,ds} \\
\vspace{0.5mm} \\
                 &=&-{\ds\frac12\int_{\pm\infty}^z\partial_s(R_0^2(s))\,ds},
\end{array}
\end{equation}
which yields the conclusion.
\qed

\vspace{3mm}
Armed with the above lemma, a statement regarding $P_{r\phi}^\pm$ and
$P_{s\phi}^\pm$ can now be made.

\begin{lemma}\label{lem:Prphi} When $\ep=0$,
\[
P_{r\phi}^\pm=-\frac12R_0',\quad P_{s\phi}^\pm=-\frac12S_0'.
\]
\end{lemma}

\proof Since $\phi'=0$ when $\ep=0$, a simple observation yields that
\[
P_{r\phi}^\pm=R_0'\phi_c^\pm,\quad P_{s\phi}^\pm=S_0'\phi_c^\pm.
\]
The conclusion now follows from the above lemma.
\qed

\begin{cor}\label{cor:Prphi} For $|z|\le M$,
\[
|P_{r\phi}^+-P_{r\phi}^-|=O(\ep),\quad
|P_{s\phi}^+-P_{s\phi}^-|=O(\ep).
\]
\end{cor}

\begin{remark} By definition, $\tp_{14}=P_{r\phi}^+-P_{r\phi}^-$ and
$\tp_{34}=P_{s\phi}^+-P_{s\phi}^-$ in Lemma \ref{lem:bfu^T}.
\end{remark}

It is now desirable to compute $P_{rs}^\pm$.
First, note that
\[
P_{r\ep}(\xi_1,\xi_2^\pm)=P_{r\om}(\xi_1,\xi_2^\pm)=0, 
\]
and that
\[
P_{rc}(\xi_1,\xi_2^\pm)=RS.
\]
Since
\begin{equation}\label{eq:P_rs}
\begin{array}{lll}
P_{rs}'&=&-SP_{rs}+2\Phi P_{r\phi}+P_{r\om}-(\ep aS-\Phi)P_{rc} \\
{}&{}&\quad\quad-2\ep a[(b+a\om)+(d_1-a)R^2+(d_2-a\al)R^4]P_{r\ep},
\end{array}
\end{equation}
upon substitution of the above relations one sees that
\[
(P_{rs}^\pm)'=-SP_{rs}^\pm+2\Phi P_{r\phi}^\pm-(\ep aS-\Phi)RS.
\]
The solution to this equation is given by
\begin{equation}\label{eq:Prspm}
R(z)P_{rs}^\pm(z)=-\ep a\int_{\pm\infty}^zR^2(s)S^2(s)\,ds+
   \int_{\pm\infty}^z R(s)\Phi(s)(R(s)S(s)+2P_{r\phi}^\pm(s))\,ds.
\end{equation}
Using Lemma \ref{lem:Prphi} and the fact that $R'=RS$ yields that
for bounded $z$,
\[
R(z)S(z)+2P_{r\phi}^\pm(z)=O(\ep),
\]
which implies, since $\Phi=O(\ep)$, that the second integral is $O(\ep^2)$.
The above argument gives the following lemma.

\begin{lemma}\label{lem:Prs} Let $M>0$ be given.
Then
\[
\begin{array}{llll}
R_0(z)P_{rs}^-(z)&=&{\ds-\ep a\int_{-\infty}^z(R'_0)^2(s)\,ds+O(\ep^2)},\quad
     &z\in(-\infty,M] \\
\vspace{.1mm} \\
R_0(z)P_{rs}^+(z)&=&{\ds\ep a\int_z^{\infty}(R'_0)^2(s)\,ds+O(\ep^2)},\quad
     &z\in[-M,\infty).
\end{array}
\]
\end{lemma}
 
\proof The conclusion follows immediately from (\ref{eq:Prspm}), taking
asymptotic expansions for $R$ and $S$, and using the fact that $R'=RS$.
\qed

\vspace{3mm}
From this lemma one can derive the following three corollaries.

\begin{cor}\label{cor:Prs} Let $M>0$ be given.
Then for $|z|\le M$,
\[
R_0(z)\tp_{rs}(z)=-\La_2'a\ep+O(\ep^2),
\]
where $\tp_{rs}=P^-_{rs}-P^+_{rs}$.
\end{cor}

\begin{cor}\label{cor:r_c} The quantity $\partial_cr^u(0)$ has the
asymptotic expansion
\[
\partial_cr^u(0)=Na\ep+O(\ep^2),
\]
where $N<0$ is given by
\[
N=\frac{\Lambda_2'}{2R_0''(0)}.
\]
\end{cor}

\proof Given the result of Lemma \ref{lem:Prs}, the conclusion 
follows immediately from the facts that
\[
P^-_{rs}(0)=-S'(0)\partial_cr^u(0),
\]
and that $S'(0)=R''(0)/R(0)$.
\qed

\vspace{3mm}
Recall the expression given for $E'''(0)$ in Lemma \ref{lem:evansthirdeval}.
Given the results of Corollary \ref{cor:Prphi} and
Lemma \ref{lem:Prs}, a definitive statement
can now be made about this quantity.

\begin{cor}\label{cor:evansthirdeval} Suppose that $E''(0)=0$.
Then
\[
E'''(0)=-(\tilde{N}a+O(e^{-\eta_4M}))\ep+O(\ep^2),
\]
where $\tilde{N}>0$ is given by
\[
\tilde{N}=3\La_2'\partial_\om\La_2.
\]
\end{cor}

\proof Substitution of the expression for $\tp_{rs}$ into the
expression for $E'''(0)$ yields that
\[
\begin{array}{lll}
E'''(0)&=&{\ds-6\ep a\La_2'\int_{-\infty}^\infty R_0(s)\partial_\om R_0(s)\,ds
              +O(e^{-\eta_4M})\ep+O(\ep^2)} \\
\vspace{.1mm} \\
  &=&{\ds-(3\La_2'\partial_\om\La_2a+O(e^{-\eta_4M}))\ep+O(\ep^2)}.
\end{array}
\]
The fact that the constant $\tilde{N}$ is positive follows immediately from
Corollary \ref{cor:Lambdaderiv}.
\qed

\section{Final Arguments}
\setcounter{equation}{0}

By Lemma \ref{lem:evanspolar}, Lemma \ref{lem:phi_om}, and 
Corollary \ref{cor:r_c} it can be seen that
\begin{equation}\label{eq:evanstwoexp}
E''(0)=C_1a(b-C_2(d_2-\frac13\al a))\ep^2+O(\ep^3),
\end{equation}
where
\[
C_1=2\La_2'\La_2\partial_\om\La_{24}<0
\]
and 
\[
C_2=-\frac{\partial_\om\La_{d_2}}{\partial_\om\La_{24}}>0.
\]
Furthermore, when $E''(0)=0$, by Corollary \ref{cor:evansthirdeval}
\begin{equation}\label{eq:evansthreeexp}
E'''(0)=-C_3a\ep+O(\ep^2),
\end{equation}
where $C_3>0$.

The proof of Theorem \ref{thm:main} is now essentially complete.
The result follows immediately from the expansions given in 
equations (\ref{eq:evanstwoexp}) and (\ref{eq:evansthreeexp}), and
the fact that $E^{(4)}(0)<0$ (Corollary \ref{cor:evansfour}).
The reason the eigenvalues are $O(\ep)$ and real follows immediately from
the fact that $E''(0)=O(\ep^2)$, while $E'''(0)=O(\ep)$.

The conclusion of Theorem \ref{thm:multipulse} follows immediately from
the work of Kapitula and Maier-Paape \cite{kapitula:sdo96}.
In order to use that work to conclude the existence of multiple 
pulse orbits, all that is necessary is to show that
$\partial_\om\phi^u(0)\neq0$.
This condition is met as a consequence of Corollary \ref{lem:defb^*}.
The fact that the multiple pulse solutions are unstable for 
$b<b^*$ follows immediately from the fact that the primary pulse
is unstable for $b<b^*$ (Alexander and Jones \cite{alexander:esa94}).
The minimal number of unstable eigenvalues also follows from that work.

\appendix
\renewcommand{\theequation}{\Alph{section}.\arabic{equation}}
\newpage

\section{Appendix: Evaluation of Constants}
\setcounter{equation}{0}

In the following,
\[
\be=-\frac{16}3\al\om.
\]

\begin{prop}\label{prop:Lambdaexp} The quantities $\Lambda_2$ and 
$\La_4$ satisfy
\[
\begin{array}{llll}
1.\quad&\Lambda_2&=&{\ds4\sqrt{\om}\frac1{\sqrt{\be}}\atanh(\sqrt{\be})} \\
\vspace{.1mm} \\
2.\quad&\Lambda_{4}&=&{\ds-16\om\frac1{\be}(\sqrt{\om}-\frac14\Lambda_2)}.
\end{array}
\]
\end{prop}

\proof Parts 1. and 2. follow immediately from a direct integration, and
can be verified with the help of Maple V (Release 4).
\qed

\begin{cor}\label{cor:Lambdaderiv} The functions $\Lambda_2$ and
$\Lambda_4$ satisfy
\[
\begin{array}{llll}
1.\quad&\partial_\om\Lambda_2&=&{\ds\frac{2}{1-\be}\om^{-1/2}} \\
\vspace{.1mm} \\
2.\quad&\partial_\om\Lambda_4&=&{\ds4\om\partial_\om\Lambda_2}.
\end{array}
\]
\end{cor}

Using the fact that
\[
\atanh(x)=\frac12\ln\frac{1+x}{1-x},
\]
Taylor series can be generated for various quantities.

\begin{cor}\label{cor:LambdaTaylor} When $0\le\be<1$ one has the
Taylor expansions
\[
\begin{array}{lrll}
1.\,&\Lambda_2&=&{\ds 4\om^{1/2}\sum_{n=0}^\infty\frac{\be^n}{2n+1}} \\
\vspace{.1mm} \\
2.\,&\Lambda_4&=&{\ds 16\om^{3/2}\sum_{n=0}^\infty\frac{\be^n}{2n+3}} \\
\vspace{.1mm} \\
3.\,&\Lambda_{24}&=&{\ds\frac3{4\om}(1-\frac{2^2}{3\cdot5}\be
    -\frac{2^2\cdot3^2}{3\cdot5^2\cdot7}\be^2
   -\frac{2^2\cdot23}{3\cdot5^3\cdot7}\be^3+O(\be^4))} \\
\vspace{.1mm} \\
4.\,&\Lambda_{d_2}&=&{\ds\frac85\om(1+\frac{3^2}{5\cdot7}\be+
   \frac{23}{5^2\cdot7}\be^2+\frac{3\cdot1879}{5^3\cdot7^2\cdot11}\be^3
   +O(\be^4))} \\
\vspace{.1mm} \\
5.\,&{\ds\partial_\om\Lambda_{d_2}}&=&{\ds\frac85(1+2\frac{3^2}{5\cdot7}\be
  +3\frac{23}{5^2\cdot7}\be^2+4\frac{3\cdot1879}{5^3\cdot7^2\cdot11}\be^3
  +O(\be^4))} \\
\vspace{.1mm} \\
6.\,&\partial_\om\Lambda_{24}&=&{\ds-\frac3{4\om^2}(1-2\frac{2^2}{3\cdot5}\be-
   3\frac{2^2\cdot3^2}{3\cdot5^2\cdot7}\be^2
     -4\frac{2^2\cdot23}{3\cdot5^3\cdot7}\be^3+O(\be^4))} \\
\vspace{.1mm} \\
7.\,&{\ds\frac{\partial_\om\Lambda_{d_2}}{\partial_\om\Lambda_{24}}}&=&
{\ds-\frac{32}{15}\om^2(1+\frac{2\cdot11}{3\cdot7}\be+\frac{73}{3^2\cdot7}\be^2
 +\frac{2^2\cdot59\cdot2017}{3^3\cdot5^2\cdot7^2\cdot11}\be^3+O(\be^4))}.
\end{array}
\]
\end{cor}


\bibliography{../papers}
\bibliographystyle{plain}



\end{document}